\documentclass{article}
\usepackage{spconf,amsmath,graphicx,booktabs,hyperref}
\usepackage{tabularx}

\ninept 
\title{Acoustic echo cancellation with the dual-signal transformation LSTM network}
\name{Nils L. Westhausen and Bernd T. Meyer \thanks{This research was supported by the DFG (Cluster of Excellence 1077/1 Hearing4all; URL: http://hearing4all.eu). The architecture was partially developed on a GPU donated by the Nvidia GPU Grant program.}}
\address{Communication Acoustics \& Cluster of Excellence Hearing4all\\
  Carl von Ossietzky University, Oldenburg, Germany}
\begin{document}
%
\maketitle
\begin{abstract}
This paper applies the dual-signal transformation LSTM network (DTLN) to the task of real-time acoustic echo cancellation (AEC). 
The DTLN combines a short-time Fourier transformation and a learned feature representation in a stacked network approach, which enables robust information processing in the time-frequency and in the time domain, which also includes phase information. The model is only trained on 60~h of real and synthetic echo scenarios. The training setup includes multi-lingual speech, data augmentation, additional noise and reverberation to create a model that should generalize well to a large variety of real-world conditions. 
The DTLN approach produces state-of-the-art performance on clean and noisy echo conditions reducing acoustic echo and additional noise robustly. The method outperforms the AEC-Challenge baseline by 0.30 in 
terms of Mean Opinion Score (MOS).
\end{abstract}
\section{Introduction}
\label{sec:intro}
Acoustic echoes can occur in audio/video calls if a speaker's voice is played back by the near-end speaker and picked up by the near-end microphone. The resulting effect of hearing an echo of your own voice can be extremely annoying, increase the listening effort and is a pressing topic in speech research - especially with the growing importance of reliable communication solutions for remote scenarios. 
A standard approach for cancelling the echo is to estimate the impulse response from the loudspeaker to the microphone by an adaptive filter such as normalized least mean squares (NLMS) \cite{article_echo} and filter the far-end signal with the estimated impulse response. This estimated signal is subtracted from the near-end microphone signal. This approach works best when only a far-end signal is present and no near-end speech is recorded by the microphone. In the case of far-end and near-end speech, also called double talk scenario, the filter will not correctly adapt or diverge \cite{benesty2001advances}. 
In this case, double talk detectors are often used to pause the adaptation. 

Recently, deep learning and neural networks have been applied to acoustic echo cancellation with convincing results \cite{Zhang2018DeepLF,fazel2020cad,ma2020acoustic, carbajal2020joint}. 
Several approaches combine neural networks and adaptive filters in a hybrid system \cite{fazel2020cad,ma2020acoustic,carbajal2020joint}.
From the deep-learning perspective, the AEC task can be seen as a speech or audio source separation problem \cite{Zhang2018DeepLF}. 
The field of speech separation quickly progressed in recent years \cite{hershey2016deep, kolbaek2017multitalker, luo2018tasnet}. However, the models for speaker separation are often concentrating on sequence processing and not on causal real-time processing. 
Because high delays are not desirable and can increase the effort in voice communication, systems that are capable of real-time processing on a frame basis are required. 
Recurrent neural networks (RNN) such as gated recurrent units (GRU) \cite{chung2014empirical} or long short term memory (LSTM) \cite{hochreiter1997long} networks are often used for models with real-time capability. 
Because of their cell structure with gates and states, LSTMs and GRUs can model time sequences on a frame basis as required for speech signals. 
RNNs were already applied to the AEC problem in \cite{Zhang2018DeepLF,fazel2020cad,ma2020acoustic}.
The deep noise suppression challenge of Interspeech 2020 \cite{reddy2020interspeech} has shown that various architectures  can be applied to real-time signal enhancement\cite{valin2020perceptually,hu2020dccrn,westhausen2020dual}. 
To address AEC as a topic of similar relevance, the AEC Challenge was proposed \cite{sridhar2020icassp} which has the aim of providing a common set of training data and objective evaluation based on an ITU P.808 framework \cite{naderi2020open} to compare various approaches. 

In this paper, the dual-signal transformation LSTM network \cite{westhausen2020dual} is adapted for real time-echo cancellation (DTLN-aec). 
The original DTLN model was shown to be beneficial and robust for reducing noise in a real-time scenario \cite{westhausen2020dual} on anechoic, reverberant and real-live test sets. 
It combines the short-time Fourier transformation (STFT) with a learned feature representation based on 1D-Conv layer in a stacked network approach. The model is based on ratio masking in the time-frequency (TF) domain and in the learned feature domain. Due to this design choice, it can leverage information from the STFT magnitude as well as from the learned feature representation. 
Since it is unclear if this approach is beneficial for AEC, we apply the model in this context with the aim of building a straight-forward RNN based end-to-end AEC system which can be easily integrated in common signal processing chains. 
For this new application, the original model is extended by feeding the far-end signal as additional information to each model block. 
This extension is similar to the procedure pursued in \cite{Zhang2018DeepLF}, with the important difference that we use a causal LSTM instead of an acausal BLSTM.
Recent publications have shown that a well-chosen training setup and data augmentation \cite{braun2020data, isik2020poconet} are crucial for achieving high speech quality for speech enhancement. 
The second goal pursued in this study is therefore to increase AEC robustness by extensive data augmentation to cover reverberation and multilingual speech. 
\section{Methods}
\label{sec:methods}
\subsection{Problem formulation}
\begin{figure}[t]
  \centering
  \includegraphics[width=0.8\linewidth]{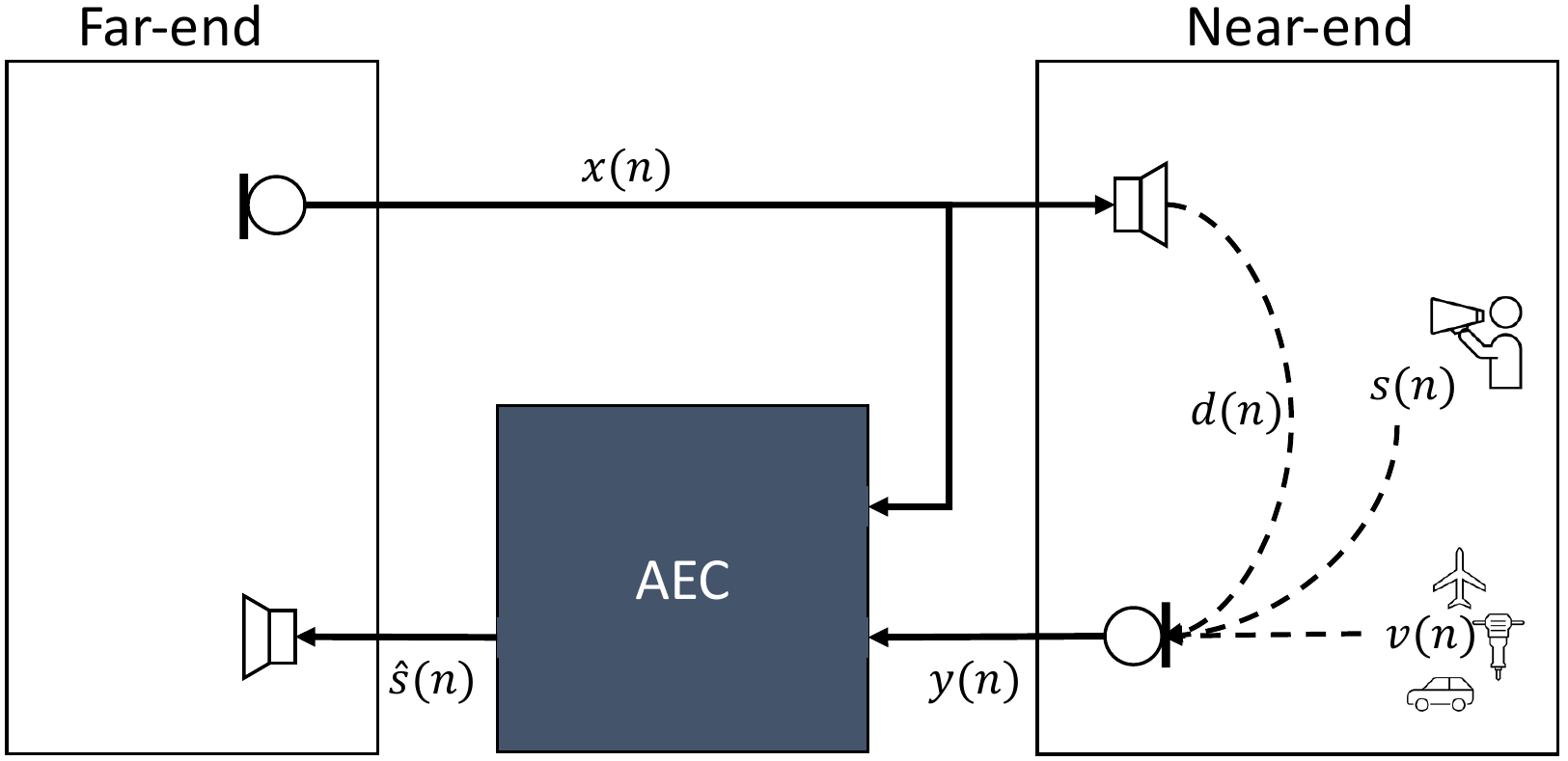}
  \caption{Illustration of an acoustic echo scenario with additional noise.}
\label{fig:aec-scenario}
\end{figure}

For an acoustic echo cancellation system two input signals are usually available, the microphone signal $y(n)$ and far-end microphone signal $x(n)$. 
The near-end microphone signal can be described as a combination of signals as following:
\begin{equation}
    y(n) = s(n) + v(n) + d(n)
\end{equation}
where $s(n)$ is the near-end speech signal, $v(n)$ is a possible near-end noise signal and $d(n)$ corresponds to the echo signal, which is a convolution of the far-end microphone signal $x(n)$ with the impulse response of the transmission path $h(n)$. 
The transmission path is a combination of a system delay created by buffering of the audio devices, the characteristics of the loudspeaker in combination with the amplifier and the transfer function between the near-end loudspeaker and the near-end microphone. 
The acoustic echo scenario is illustrated in \autoref{fig:aec-scenario}. The desired signal is the near-end speech signal $s(n)$, while all other signal parts should be removed. This task is an audio source separation task. If only far-rend and noise signals are present, the desired signal is silence. 
\subsection{DTLN model adapted for AEC}
\label{sub:network}
In the context of the DNS-Challenge at Interspeech 2020 \cite{reddy2020interspeech}, the Dual Signal Transformation LSTM network (DTLN) \cite{westhausen2020dual} was developed to reduce the noise in noisy speech mixtures. The DTLN approach was adapted to the AEC task (DTLN-aec\footnote{Pretrained model available at \url{https://github.com/breizhn/DTLN-aec}}) and is described in the following.

The network consists of two separation cores. Each separation core has two LSTM layers and a fully-connected layer with sigmoid activation to predict masks. The first separation core is fed by concatenated normalized log power spectra of the near-end and far-end microphone signal. Each microphone signal is individually normalized by instant layer normalization (iLN) to account for level variations. Instant layer normalization is similar to standard layer normalization \cite{ba2016layer}, where each frame is normalized individually but without accumulating statistics over time. This concept was introduced as channel-wise layer normalization in \cite{luo2018conv2}. The first core predicts a time frequency mask which is applied to the unnormalized magnitude STFT of the near-end microphone signal. The estimated magnitude is transformed back to the time domain with an inverse FFT using the phase of the original near-end microphone signal.

\begin{figure*}[t]
  \centering
  \includegraphics[width=0.80\linewidth]{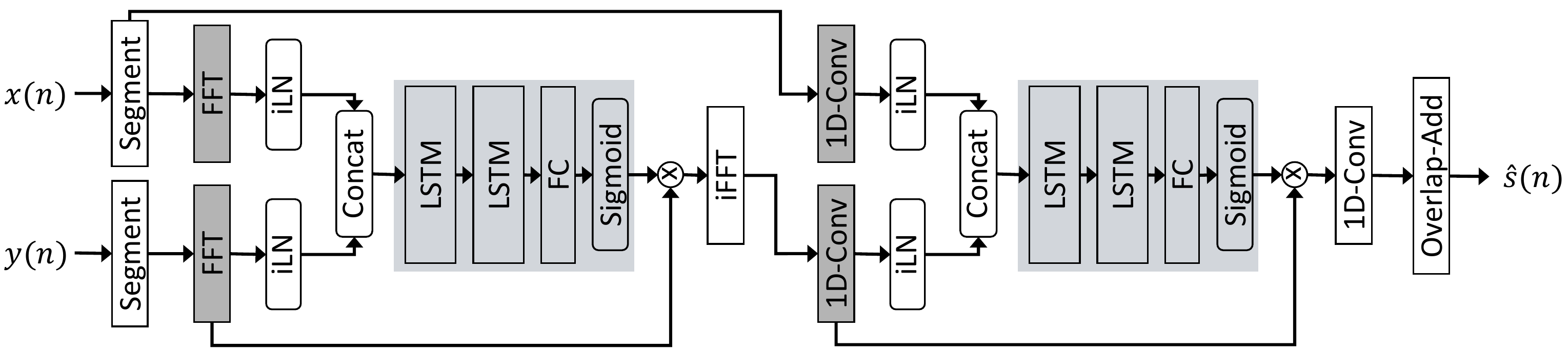}
  \caption{Illustration of the proposed DTLN-aec model architecture. The processing chain on the left shows the first separation core using the STFT signal transformation (split in segmentation and FFT for both near-end and far-end microphone signal), while the building blocks on the right represent the second core with learned feature transformations based on 1D-Conv layers applied to the output of the first core and the segmented far-end microphone signal.}
\label{fig:dns-network}
\end{figure*}

The second core uses a learned feature representation created with a 1D-Conv layer. This approach is inspired by \cite{luo2018tasnet, luo2019conv}. The core is fed with the normalized feature representation of the previously predicted signal and the normalized feature representation of the far-end microphone signal. For transforming both signals to the time domain, the same weights are applied but the normalization with iLN is performed individually to enable a separate scaling and bias for each representation. 
The predicted mask of the second core is multiplied with the unnormalized feature representation of the output of the first core. This estimated feature representation is transformed back to the time domain with a 1D-Conv layer. For reconstructing the continuous time signal an overlap-add procedure is used. The model architecture is visualized in \autoref{fig:dns-network}. 

For the task of echo cancellation, a frame length of 32~ms and a frame shift of 8~ms was chosen. The FFT size is 512 and the size of the learned feature representation is also 512. Since the removal of speech and noise from speech can be quite challenging, 512 LSTM units per layer were chosen compared to the rather small model in \cite{westhausen2020dual}. 
This results in a total of 10.3M parameters for the current model. 
Additionally, models with 128 and 256 units per layer were trained to explore how model performance scales with size.
\subsection{Datasets and dataset preparation}
Two training datasets are provided through the challenge, one with synthesized data and one with real recordings. 
The synthetic dataset was derived from the dataset created for \cite{reddy2020interspeech}. 
The dataset includes 10,000 examples containing single-talk, double-talk, near-end noise, far-end noisy and various nonlinear distortion situations, where each example contains far-end speech, echo signal, near-end speech, and the near-end microphone signal. 
The first 500 examples contain data from speakers whose data is not contained in any other test dataset. 
This dataset will be used for instrumental evaluation and is refered to as "double-talk test set". For more details, see the paper describing the AEC-Challenge \cite{sridhar2020icassp}. For training, only the far-end signals and the echo signals were used and cut into chunks of 4~s. 
The real dataset consists of different real environments with human speakers and signals captured with varying different devices. 
Detailed information on this data is provided in \cite{sridhar2020icassp}. 
As before, only the far-end signal and the echo signal are used in chunks of 4~s from this dataset. 
For the evaluation with the P. 808 framework, a blind test set was provided by the challenge organizers. The blind test set consist of approximately 800 recordings divided into a clean and noisy subset.

Clean speech from the multilingual data gathered for \cite{reddy2020icassp} were chosen 
as near-end signals. 
The dataset contains French, German, Italian, Mandarin, English, Russian and Spanish speech. 
The various sources of the original data are described in \cite{reddy2020icassp}. The German data was excluded because of its poor quality. 
The speech signals were segmented into samples with a duration of 4\,s. 
Samples with an RMS smaller or equal to zero are discarded. An RMS smaller than zero can result from rounding errors.
As an additional mechanism to exclude noisy signals, each file was processed by the speech enhancement model proposed in \cite{westhausen2020dual} to estimate a speech and a noise signal by subtracting the estimated speech signal from the noisy signal. 
The speech file is discarded if the SNR is lower than 5~dB. 
Finally, 20~h from each language are taken to create a dataset of 120~h of multilingual speech. 

To cover noise types with a high variance in the echo scenario, we used the noise corpus provided by \cite{reddy2020icassp}. 
As before, the noise files were cut into 4 second samples, and each sample with an RMS smaller or equal to zero was discarded. 
Additionally, instrumental music from the MUSAN corpus \cite{snyder2015musan} was added (again, after a 4\,s segmentation). This results in approximately 140\,h of noise. 

Finally, to build realistic echo scenarios that reflect the influence of divers amounts of reverberation, the impulse responses (IR) dataset gathered for \cite{ko2017study} were used. The dataset contains real impulse responses from various sources such as \cite{kinoshita2013reverb, nakamura2000acoustical, jeub2009binaural} and simulated ones based on the image method \cite{allen1979image}.
For each impulse response, the begin of the direct path was identified and set to position 0 as proposed in \cite{isik2020poconet}.
\subsection{Training and data augmentation}
All training samples are created online during training without using fixed combinations of near-end speech, far-end speech, noise and IRs. In total, 60~h of echo scenarios are used, 48~h for training and the remaining 12~h for training validation.
For training, all far-end and echo signals provided by the challenge organizers are used (approximately 32~h of data). 
To create additional echo data, 28~h of speech are used from the previously created multilingual dataset. Each speech file is convolved with a randomly chosen IR, and each IR is divided by the absolute value of the first sample. In the next step, all samples except for the first sample are multiplied by a gain randomly taken from a uniform distribution between -25 and 0 to augment the IRs. This procedure is again inspired by \cite{isik2020poconet}.

In 50\% of the cases, a noise sample is added with an SNR randomly taken from a normal distribution with a mean 5~dB and standard deviation 10~dB to account for a noisy far-end signal. For creating the echo signal, the previously created far-end signal is delayed by a random value between 10 and 100~ms to simulate a processing and transmission delay. The delayed signal is filtered by a band-pass signal with a random lower cut-off frequency between 100 and 400~Hz and a higher cut-off frequency between 6000 and 7500~Hz. This step introduces additional variance and models the often poor acoustic transmission characteristics of in-device loudspeakers especially in the low-frequency region. The echo signal is finally convolved with same IR as the near-end signal. Additional non-linearities are not included since the original challenge data set already covers this aspect.

For the near-end signals, 60\,h from the multilingual data set are used. Each speech file is convolved by randomly selected IR, which is randomly scaled as explained for the synthetic far-end signals. Random spectral shaping as suggested by \cite{braun2020data} for noise reduction is applied to the speech signal to increase robustness and model various transmission effects.

In 70\% of the cases, noise is added to the near-end speech with a an SNR taken from a normal distribution with mean 5 and standard deviation 10 to shift the focus to the more challenging noisy near-end condition. Random spectral shaping is also applied to the noise signal independently. 

In 5\% of the cases, a near-end speech segment of random duration is discarded to account for far-end-only scenarios. 
In 90\% of the cases, the echo signal is added to the near-end speech with a speech-to-echo ratio taken from a normal distribution with a 0\,dB mean and standard deviation of 10\,dB.
The echo signal as well as the far-end speech signal is applied with random spectral shaping. If no echo is applied, the far-end signal is set to zero or to low-level noise in the range between -70 and -120~dB RMS with random spectral shaping.
All signals used as input to the model are subject to a random gain chosen from a uniform distribution ranging from -25 to 0~dB relative to the clipping point. 

The SNR-loss in time domain as first proposed in \cite{kavalerov2019universal} was chosen as cost function. 
The SNR-loss is scale-dependent, which is desirable for real-time applications and implicitly integrates phase information because it is calculated in time domain.
The model is trained with the Adam optimizer \cite{Kingma2015AdamAM} for 100 epochs with an initial learning rate of 2e-4 for 512 LSTM units, 5e-4 for 256 units and 1e-3 for 128 units. The learning rate is multiplied by 0.98 every two epochs. Gradient norm clipping with a value of 3 was applied. The batch size was set to 16 and the sample length to 4\,s. 
Between consecutive LSTM layers, 25\% of dropout was introduced to reduce overfitting. 
The model was evaluated every epoch using the validation set. 
The model with the best performance on the validation set was used for testing. 
\subsection{Baseline systems}
The challenge organizers also provide a baseline which is based on \cite{xia2020weighted}. The baseline consist of two GRU layers and a fully-connected network with sigmoid activation to predict a time-frequency mask. 
The model is fed with the concatenated short-time log-power-spectra of the microphone and the loop-back signal and predicts a spectral suppression mask which is applied to the STFT magnitude of the microphone signal. The predicted magnitude spectra are transformed back to the time domain with an inverse STFT using the phase of the microphone signal.
Since the baseline model was not accessible within the challenge, an additional baseline system was trained to quantify the performance of a stacked network compared to a model with consecutive LSTM layers using time-frequency masking. 
The model has four consecutive LSTM layers with 512 units each, followed by a fully-connected part with sigmoid activation to predict the TF-mask. 
The input to the model equals the first separation core of the DTLN-aec model. The mask is multiplied with the unormalized magnitude of the near-end microphone signal and transformed back to the time domain. This configuration results in a model with 8.5M parameters. The model is trained with the same setup as the DTLN-aec model.
\begin{table*}[t]
    \caption{Results in terms of PESQ [MOS] and SI-SDR [dB] on the clean, noisy far-end signal, noisy near-end signal and noisy far- and near-end signal subsets of the double-talk test set.}
  \centering
  \begin{tabular}{l r r r r r r r r r r r r r}
  \toprule
  & &  \multicolumn{2}{c}{\textbf{clean}} & & \multicolumn{2}{c}{\textbf{far-end noisy}} &  & \multicolumn{2}{c}{\textbf{near-end noisy}} &  & \multicolumn{2}{c}{\textbf{both noisy}}
        \\
    \midrule
    \textbf{Method} & \textbf{\# Params/Units} &\textbf{PESQ} &  \textbf{SI-SDR}   &
    & \textbf{PESQ} & \textbf{SI-SDR}   &
    & \textbf{PESQ} & \textbf{SI-SDR}   &
    & \textbf{PESQ} & \textbf{SI-SDR}  \\
    \midrule
     Unprocessed &  &  2.05 & 0.01  &  
     & 1.95 & -0.88  & 
     & 1.77 & -1.58  & 
     & 1.68 & -2.35  \\
     \midrule
     Baseline & 8.5M / 512  &  2.66 & 13.02 &  
     & 2.55 & 12.20  & 
     & 2.34 & 11.01  & 
     & 2.31 & 10.29  \\
   
   DTLN-aec & 1.8M / 128 & 2.57 & 12.00 &
                         & 2.49 & 11.66 &
                         & 2.29 & 10.81 &
                         & 2.27 & 10.09 \\
   DTLN-aec & 3.9M / 256 & 2.68 & 13.34 &
                         & 2.60 & 12.65 &
                         & 2.38 & 11.69 &
                         & 2.34 & 11.01 \\
    \midrule
    \textbf{DTLN-aec} & 10.4M / 512 & \textbf{2.81} & \textbf{14.15} &
             & \textbf{2.75} & \textbf{13.59} &
             & \textbf{2.53} & \textbf{12.59} &
             & \textbf{2.46} & \textbf{11.83} \\
     \bottomrule
  \end{tabular}
  \label{tab:results}
\end{table*}
\subsection{Objective and subjective evaluation}
\label{sub:Eval}
The widely used PESQ \cite{pesq} and ERLE \cite{erle} measures for evaluating AEC systems are often not correlating well with  subjective ratings \cite{sridhar2020icassp}. Nevertheless objective measures can be an indication if models are performing as intended. Because the dataset used for instrumental evaluation contains only double talk scenarios and because the AEC problem is seen as a source-separation problem the SI-SDR \cite{le2019sdr} is used to evaluate the separation performance. Additionally PESQ is used for an indication of speech quality.
The measures are used to compare the additional baseline and the differently sized DTLN-aec models on double-talk test set.

To get a better impression on the real AEC performance the challenge organizers conducted a study based on the ITU P.808 crowd-sourcing framework \cite{naderi2020open} on the Amazon Mechanical Turk platform. There are in total four scenarios evaluated: single-talk near-end (P.808), single-talk far-end (P.831 \cite{echo_subj}), double-talk echo (P.831) and double-talk other disturbances (P.831). For more details on the rating process see \cite{sridhar2020icassp}.
\section{Results}
\label{sec:Results}
The results of the objective and subjective evaluations are shown in \autoref{tab:results} and in \autoref{tab:mos_ver1}, respectively.
\\
\textbf{Objective results}:
For all conditions all models are showing an improvement over the unprocessed condition. The largest improvement is observed for the DTLN-aec with 512 units and the lowest for the DTLN-aec with 128 units. The baseline is outperformed by the models with 256 and 512 units. The improvement relative to the unprocessed condition in terms of PESQ and SI-SDR are relatively stable over all noise conditions for all models. The mean SI-SDR improvement for the model with 512 units over all conditions is 14.24~dB and the mean PESQ improvement is 0.78~MOS.
\\
\textbf{Subjective results}:
In all conditions, except for the clean single-talk near-end condition, the DTLN-aec model outperforms the AEC-Challenge baseline. The mean improvement in terms of MOS is 0.34 and 0.26 for the clean and noisy subset, respectively. 
\begin{table}[t]
  
  \caption{Subjective ratings in terms of MOS for the blind test set of the AEC-Challenge. The confidence interval is 0.02 for the clean and noisy subset (ST = single talk, DT = double talk, NE = near-end, FE = far-end). }
  \centering
  \begin{tabularx}{0.475\textwidth}{ l X X X X X X X X}
    \toprule
      \textbf{Method} & \multicolumn{2}{c}{\textbf{ST-NE}}  &\multicolumn{2}{c}{ \textbf{ST-FE} }  & \multicolumn{2}{c}{\textbf{DT-Echo}} & \multicolumn{2}{c}{\textbf{DT-other} }
        \\
     
    \midrule
   
     & clean & noisy  & clean & noisy  & clean & noisy  & clean & noisy \\
    \midrule
     Baseline & \textbf{3.99} & 3.58 & 4.09 & 3.58 & 3.89 & 3.78 & 3.33 & 3.23 \\
     DTLN-aec & 3.98 & \textbf{3.68} & \textbf{4.46} & \textbf{3.83} & \textbf{4.34} & \textbf{4.00} & \textbf{3.86} & \textbf{3.68}\\
    
    \bottomrule
  \end{tabularx}
  \label{tab:mos_ver1}
\end{table}
\\
\textbf{Results on execution time}:
To comply with the rules of the AEC-Challenge, the execution time for one audio frame must be less than the frame-shift, in our case 8~ms. 
The execution time was measured on two CPUs with a TensorFlow lite model of the DTLN-aec with 512 LSTM units per layer. 
We measured execution times of 3.06~ms (using dual-core I5-3320M at 2.6~GHz CPU) and 0.97~ms (with an I5-6600K quad-core CPU clocked at 3.5~GHz), both of which comply with AEC-Challenge rules.
\section{Discussion}
\label{sec:discussion}
When comparing the models in different sizes, the DTLN-aec model seems to scale well with respect to the number of parameters: The small model with 128 already reaches a good improvement over the noisy condition, the model with 256 units outperforms the baseline with less than half its parameters. This also shows the advantage of using a stacked model compared to models with four consecutive LSTM layers.
For the AEC task it can be an advantage to use a model with higher modeling capacity since it is not only separating speech from noise, but separating speech from speech, which can be a more challenging task - especially when voices have similar characteristics. 
For applications tailored to specific hardware, the size of the model could be chosen depending on constraints such as computational resources and power consumption. 

All models including the four-layer baseline are showing a constant improvement over the unprocessed signals for the double-talk test set. 
This suggests that the training setup is able to represent the variance of the  four tested double-talk conditions. 
The same conclusion is supported by the results on the blind test set. 
The model shows an improvement over the AEC-challenge baseline in all conditions containing an echo signal or/and noise. 
The training set only contains English speech samples, so the generalization over multiple languages was not evaluated in our study, which should be addressed in the future.
The result on the clean ST-NE condition only shows that the baseline and the DTLN-aec model have similar impact on clean near-end speech without noise and echo, 
and their detrimental effect on the optimal signals is very limited. 
Nevertheless, when listening to the processed signals, some residual noise is still audible in some conditions. In a future improvement of the DTLN-aec model, an additional noise reduction to further increase the speech quality could be added. To reduce residual noise in far-end only conditions, a voice activity detection could be added for detecting near-end speech and gating the signal in the absence of near-end speech.
\section{Conclusion}
\label{sec:conclusion}
This study has shown that the dual-signal transformation LSTM network (DTLN-aec) can successfully be applied to real-time acoustic echo cancellation. 
DTLN-aec produced state-of-the-art performance on the blind test-set of the AEC-Challenge and synthetic double-talk test set and 
is among the top five models in the AEC-Challenge. The model was trained with extensive data augmentation on publicly available data, which results in a reproducible and robust model for real-world applications. 
\vfill\pagebreak
%
%
\bibliographystyle{IEEEbib}
\bibliography{strings,refs}

\begin{thebibliography}{10}

\bibitem{article_echo}
Gerald Enzner, Herbert Buchner, Alexis Favrot, and Fabian Kuech,
\newblock ``Chapter 30. acoustic echo control,''
\newblock {\em Academic Press Library in Signal Processing}, vol. 4, 12 2014.

\bibitem{benesty2001advances}
Jacob Benesty, Tomas G{\"a}nsler, Dennis~R Morgan, M~Mohan Sondhi, Steven~L
  Gay, et~al.,
\newblock {\em Advances in network and acoustic echo cancellation},
\newblock Springer, 2001.

\bibitem{Zhang2018DeepLF}
H.~Zhang and D.~Wang,
\newblock ``Deep learning for acoustic echo cancellation in noisy and
  double-talk scenarios,''
\newblock in {\em INTERSPEECH}, 2018.

\bibitem{fazel2020cad}
Amin Fazel, Mostafa El-Khamy, and Jungwon Lee,
\newblock ``Cad-aec: Context-aware deep acoustic echo cancellation,''
\newblock in {\em ICASSP 2020-2020 IEEE International Conference on Acoustics,
  Speech and Signal Processing (ICASSP)}. IEEE, 2020, pp. 6919--6923.

\bibitem{ma2020acoustic}
Lu~Ma, Hua Huang, Pei Zhao, and Tengrong Su,
\newblock ``Acoustic echo cancellation by combining adaptive digital filter and
  recurrent neural network,''
\newblock {\em arXiv preprint arXiv:2005.09237}, 2020.

\bibitem{carbajal2020joint}
Guillaume Carbajal, Romain Serizel, Emmanuel Vincent, and Eric Humbert,
\newblock ``Joint nn-supported multichannel reduction of acoustic echo,
  reverberation and noise,''
\newblock {\em IEEE/ACM Transactions on Audio, Speech, and Language
  Processing}, vol. 28, pp. 2158--2173, 2020.

\bibitem{hershey2016deep}
John~R Hershey, Zhuo Chen, Jonathan Le~Roux, and Shinji Watanabe,
\newblock ``Deep clustering: Discriminative embeddings for segmentation and
  separation,''
\newblock in {\em 2016 IEEE International Conference on Acoustics, Speech and
  Signal Processing (ICASSP)}. IEEE, 2016, pp. 31--35.

\bibitem{kolbaek2017multitalker}
Morten Kolb{\ae}k, Dong Yu, Zheng-Hua Tan, and Jesper Jensen,
\newblock ``Multitalker speech separation with utterance-level permutation
  invariant training of deep recurrent neural networks,''
\newblock {\em IEEE/ACM Transactions on Audio, Speech, and Language
  Processing}, vol. 25, no. 10, pp. 1901--1913, 2017.

\bibitem{luo2018tasnet}
Yi~Luo and Nima Mesgarani,
\newblock ``Tasnet: time-domain audio separation network for real-time,
  single-channel speech separation,''
\newblock in {\em 2018 IEEE International Conference on Acoustics, Speech and
  Signal Processing (ICASSP)}. IEEE, 2018, pp. 696--700.

\bibitem{chung2014empirical}
Junyoung Chung, Caglar Gulcehre, Kyunghyun Cho, and Yoshua Bengio,
\newblock ``Empirical evaluation of gated recurrent neural networks on sequence
  modeling,''
\newblock in {\em NIPS 2014 Workshop on Deep Learning, December 2014}, 2014.

\bibitem{hochreiter1997long}
Sepp Hochreiter and J{\"u}rgen Schmidhuber,
\newblock ``Long short-term memory,''
\newblock {\em Neural computation}, vol. 9, no. 8, pp. 1735--1780, 1997.

\bibitem{reddy2020interspeech}
Chandan~KA Reddy, Vishak Gopal, Ross Cutler, Ebrahim Beyrami, Roger Cheng,
  Harishchandra Dubey, Sergiy Matusevych, Robert Aichner, Ashkan Aazami,
  Sebastian Braun, et~al.,
\newblock ``The interspeech 2020 deep noise suppression challenge: Datasets,
  subjective testing framework, and challenge results,''
\newblock {\em arXiv preprint arXiv:2005.13981}, 2020.

\bibitem{valin2020perceptually}
Jean-Marc Valin, Umut Isik, Neerad Phansalkar, Ritwik Giri, Karim Helwani, and
  Arvindh Krishnaswamy,
\newblock ``A perceptually-motivated approach for low-complexity, real-time
  enhancement of fullband speech,''
\newblock {\em arXiv preprint arXiv:2008.04259}, 2020.

\bibitem{hu2020dccrn}
Yanxin Hu, Yun Liu, Shubo Lv, Mengtao Xing, Shimin Zhang, Yihui Fu, Jian Wu,
  Bihong Zhang, and Lei Xie,
\newblock ``Dccrn: Deep complex convolution recurrent network for phase-aware
  speech enhancement,''
\newblock {\em arXiv preprint arXiv:2008.00264}, 2020.

\bibitem{westhausen2020dual}
Nils~L Westhausen and Bernd~T Meyer,
\newblock ``Dual-signal transformation lstm network for real-time noise
  suppression,''
\newblock {\em arXiv preprint arXiv:2005.07551}, 2020.

\bibitem{sridhar2020icassp}
Kusha Sridhar, Ross Cutler, Ando Saabas, Tanel Parnamaa, Hannes Gamper,
  Sebastian Braun, Robert Aichner, and Sriram Srinivasan,
\newblock ``Icassp 2021 acoustic echo cancellation challenge: Datasets and
  testing framework,''
\newblock {\em arXiv preprint arXiv:2009.04972}, 2020.

\bibitem{naderi2020open}
Babak Naderi and Ross Cutler,
\newblock ``An open source implementation of itu-t recommendation p. 808 with
  validation,''
\newblock {\em arXiv preprint arXiv:2005.08138}, 2020.

\bibitem{braun2020data}
Sebastian Braun and Ivan Tashev,
\newblock ``Data augmentation and loss normalization for deep noise
  suppression,''
\newblock in {\em International Conference on Speech and Computer}. Springer,
  2020, pp. 79--86.

\bibitem{isik2020poconet}
Umut Isik, Ritwik Giri, Neerad Phansalkar, Jean-Marc Valin, Karim Helwani, and
  Arvindh Krishnaswamy,
\newblock ``Poconet: Better speech enhancement with frequency-positional
  embeddings, semi-supervised conversational data, and biased loss,''
\newblock {\em arXiv preprint arXiv:2008.04470}, 2020.

\bibitem{ba2016layer}
Jimmy~Lei Ba, Jamie~Ryan Kiros, and Geoffrey~E Hinton,
\newblock ``Layer normalization,''
\newblock {\em arXiv preprint arXiv:1607.06450}, 2016.

\bibitem{luo2018conv2}
Yi~Luo and Nima Mesgarani,
\newblock ``Conv-tasnet: Surpassing ideal time-frequency magnitude masking for
  speech separation,''
\newblock {\em arXiv preprint arXiv:1809.07454}, 2018.

\bibitem{luo2019conv}
Yi~Luo and Nima Mesgarani,
\newblock ``Conv-tasnet: Surpassing ideal time--frequency magnitude masking for
  speech separation,''
\newblock {\em IEEE/ACM transactions on audio, speech, and language
  processing}, vol. 27, no. 8, pp. 1256--1266, 2019.

\bibitem{reddy2020icassp}
Chandan~KA Reddy, Harishchandra Dubey, Vishak Gopal, Ross Cutler, Sebastian
  Braun, Hannes Gamper, Robert Aichner, and Sriram Srinivasan,
\newblock ``Icassp 2021 deep noise suppression challenge,''
\newblock {\em arXiv preprint arXiv:2009.06122}, 2020.

\bibitem{snyder2015musan}
David Snyder, Guoguo Chen, and Daniel Povey,
\newblock ``Musan: A music, speech, and noise corpus,''
\newblock {\em arXiv preprint arXiv:1510.08484}, 2015.

\bibitem{ko2017study}
Tom Ko, Vijayaditya Peddinti, Daniel Povey, Michael~L Seltzer, and Sanjeev
  Khudanpur,
\newblock ``A study on data augmentation of reverberant speech for robust
  speech recognition,''
\newblock in {\em 2017 IEEE International Conference on Acoustics, Speech and
  Signal Processing (ICASSP)}. IEEE, 2017, pp. 5220--5224.

\bibitem{kinoshita2013reverb}
Keisuke Kinoshita, Marc Delcroix, Takuya Yoshioka, Tomohiro Nakatani, Emanuel
  Habets, Reinhold Haeb-Umbach, Volker Leutnant, Armin Sehr, Walter Kellermann,
  Roland Maas, et~al.,
\newblock ``The reverb challenge: A common evaluation framework for
  dereverberation and recognition of reverberant speech,''
\newblock in {\em 2013 IEEE Workshop on Applications of Signal Processing to
  Audio and Acoustics}. IEEE, 2013, pp. 1--4.

\bibitem{nakamura2000acoustical}
Satoshi Nakamura, Kazuo Hiyane, Futoshi Asano, Takanobu Nishiura, and Takeshi
  Yamada,
\newblock ``Acoustical sound database in real environments for sound scene
  understanding and hands-free speech recognition,''
\newblock {\em LREC}, 2000.

\bibitem{jeub2009binaural}
Marco Jeub, Magnus Schafer, and Peter Vary,
\newblock ``A binaural room impulse response database for the evaluation of
  dereverberation algorithms,''
\newblock in {\em 2009 16th International Conference on Digital Signal
  Processing}. IEEE, 2009, pp. 1--5.

\bibitem{allen1979image}
Jont~B Allen and David~A Berkley,
\newblock ``Image method for efficiently simulating small-room acoustics,''
\newblock {\em The Journal of the Acoustical Society of America}, vol. 65, no.
  4, pp. 943--950, 1979.

\bibitem{kavalerov2019universal}
Ilya Kavalerov, Scott Wisdom, Hakan Erdogan, Brian Patton, Kevin Wilson,
  Jonathan Le~Roux, and John~R Hershey,
\newblock ``Universal sound separation,''
\newblock in {\em 2019 IEEE Workshop on Applications of Signal Processing to
  Audio and Acoustics (WASPAA)}. IEEE, 2019, pp. 175--179.

\bibitem{Kingma2015AdamAM}
Diederik~P. Kingma and Jimmy Ba,
\newblock ``Adam: A method for stochastic optimization,''
\newblock {\em CoRR}, vol. abs/1412.6980, 2015.

\bibitem{xia2020weighted}
Yangyang Xia, Sebastian Braun, Chandan~KA Reddy, Harishchandra Dubey, Ross
  Cutler, and Ivan Tashev,
\newblock ``Weighted speech distortion losses for neural-network-based
  real-time speech enhancement,''
\newblock in {\em ICASSP 2020-2020 IEEE International Conference on Acoustics,
  Speech and Signal Processing (ICASSP)}. IEEE, 2020, pp. 871--875.

\bibitem{pesq}
``{ITU-T P.862: Perceptual evaluation of speech quality (PESQ): An objective
  method for end-to-end speech quality assessment of narrow-band telephone
  networks and speech codecs.},'' 2001.

\bibitem{erle}
``{ITU-T G.168: Digital network echo cancellers.},'' 2012.

\bibitem{le2019sdr}
Jonathan Le~Roux, Scott Wisdom, Hakan Erdogan, and John~R Hershey,
\newblock ``Sdr--half-baked or well done?,''
\newblock in {\em ICASSP 2019-2019 IEEE International Conference on Acoustics,
  Speech and Signal Processing (ICASSP)}. IEEE, 2019, pp. 626--630.

\bibitem{echo_subj}
``{ITU-T P.831: Subjective performance evaluation of network echo
  cancellers.},'' 1998.

\end{thebibliography}

\end{document}